\documentclass[a4paper]{jpconf}
\usepackage{graphicx}
\usepackage{hyperref}
\usepackage{cite}
\usepackage{lineno}

\begin{document}
\title{Parallelized JUNO simulation software based on SNiPER}

\newcommand{\IHEP}{$^1$}
\newcommand{\SDU}{$^2$}
\newcommand{\SYSU}{$^3$}

\author{Tao Lin\IHEP, Jiaheng Zou\IHEP, Weidong Li\IHEP, Ziyan Deng\IHEP, Guofu Cao\IHEP, Xingtao Huang\SDU{} and Zhengyun You\SYSU{}}
\author{(On Behalf of the JUNO Collaboration)}
\address{\IHEP Institute of High Energy Physics, Chinese Academy of Sciences, Beijing, China}
\address{\SDU Shandong University, Jinan, China}
\address{\SYSU Sun Yat-sen University, Guangzhou, China}

\ead{lintao@ihep.ac.cn, zoujh@ihep.ac.cn}


\begin{abstract}
The Jiangmen Underground Neutrino Observatory (JUNO) is a neutrino experiment to determine neutrino mass hierarchy. It has a central detector used for neutrino detection, which consists of a spherical acrylic vessel containing 20 kt liquid scintillator (LS) and about 18,000 20-inch photomultiplier tubes (PMTs) to collect light from LS. Around the central detector, there is a water pool to shield radioactivities. The outer water pool is also equipped with about 2000 PMTs to measure cosmic ray muons by detecting Cherenkov light.

As one of the important parts in JUNO offline software, the serial simulation framework is developed based on SNiPER. It is in charge of physics generator, detector simulation, event mixing and digitization. However Geant4 based detector simulation of such a large detector is time-consuming and challenging. It is necessary to take full advantages of parallel computing to speedup simulation. Starting from version 10.0, Geant4 supports event-level parallelism. Even though based on pthread, it could be extended with other libraries such as Intel TBB and MPI. Therefore it is possible to parallelize JUNO simulation framework via integrating Geant4 and SNiPER.

In this paper, our progress in developing parallelized simulation software are presented. The SNiPER framework can run in sequential mode, Intel TBB mode or other modes. The SNiPER task component is in charge of event loop, which is like a simplified application manager. Two types of tasks are introduced in the simulation framework, one is global task and another is worker task. The global task will run only once to initialize detector geometry and physics processes before any other tasks spawned. Later it is accessed by other tasks passively. The worker tasks will be spawned after global task is done. In each worker task, a Geant4 run manager is invoked to do the real simulation. Therefore the simulation framework and the underlying TBB have been decoupled. Finally, the software performance of parallelized JUNO simulation software is also presented. 
\end{abstract}

\section{Introduction}
\begin{figure}[h]
\begin{center}
\includegraphics[width=20pc]{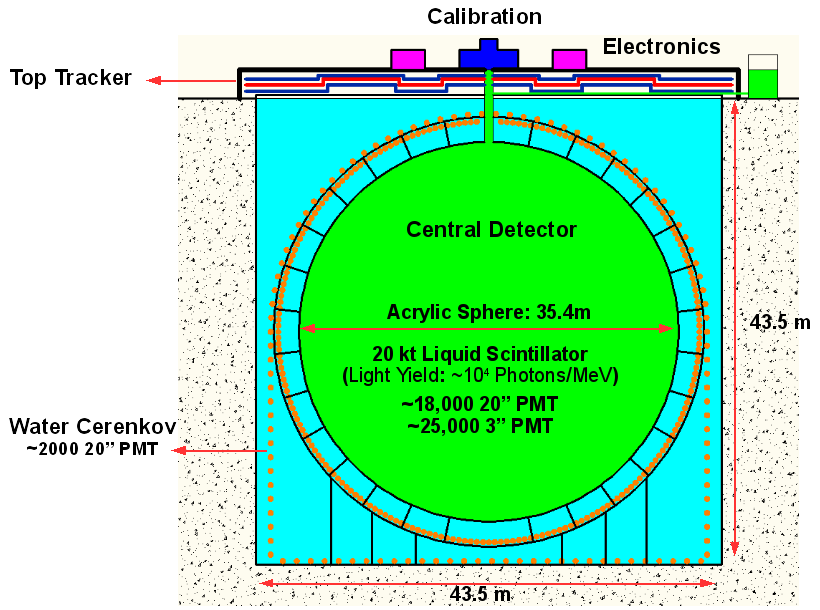}\hspace{2pc}%
\caption{\label{fig:10}Schematic view of JUNO detector}
\end{center}
\end{figure}

JUNO \cite{An:2015jdp,Djurcic:2015vqa} is an underground neutrino experiment to measure neutrino mass hierarchy and oscillation parameters. It will be located in southern China, about 53~km away from two nuclear power plants. Figure~\ref{fig:10} shows the schematic view of the JUNO detector. To detect neutrinos, the innermost part, which is called central detector, is filled with 20~kt liquid scintillator (LS). When neutrinos react with LS, lights are yielded and then collected by the surrounding photomultiplier tubes (PMTs). To suppress the radioactivities from PMTs and rocks, water is filled around the spherical acrylic vessel. The outer water pool is also equipped with PMTs to veto cosmic ray muon events by detecting Cherenkov light. On the top of the water pool, a top tracker is used to measure muons.

\begin{figure}[h]
\begin{center}
\includegraphics[width=30pc]{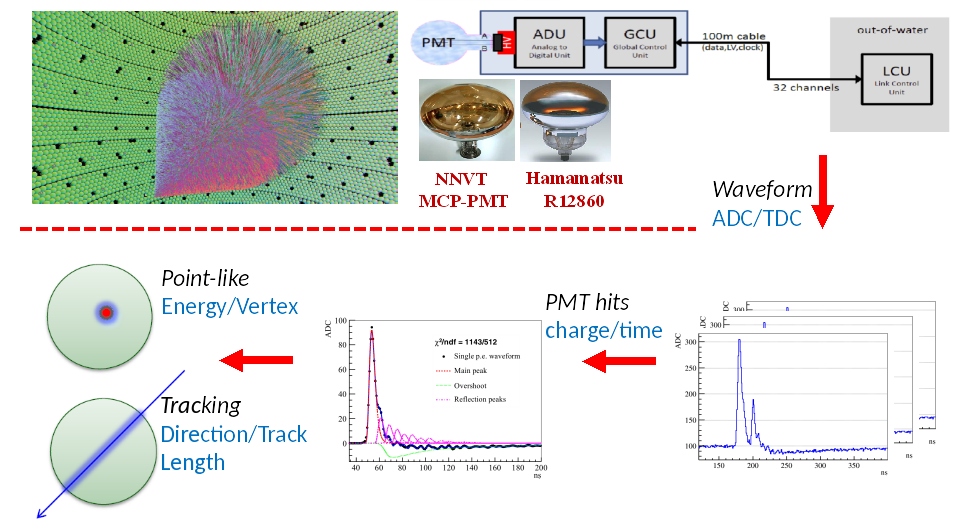}\hspace{2pc}%
\caption{\label{fig:11}Offline data processing chain}
\end{center}
\end{figure}

In the JUNO experiment, offline software is an important part. Simulation software is used for detector design and optimization, algorithms tuning and physics studies. PMT waveform reconstruction is used to calibrate charge and time for each PMT. Reconstruction algorithms are used to reconstruct events and get physics objects for physics analysis. These algorithms are built upon a software framework called SNiPER \cite{Zou:2015ioy}, and compose a full chain of data processing \cite{Huang:junooffline}, as shown in Figure~\ref{fig:11}.

\section{Towards parallelized detector simulation framework}
\begin{figure}[h]
\begin{center}
\includegraphics[width=20pc]{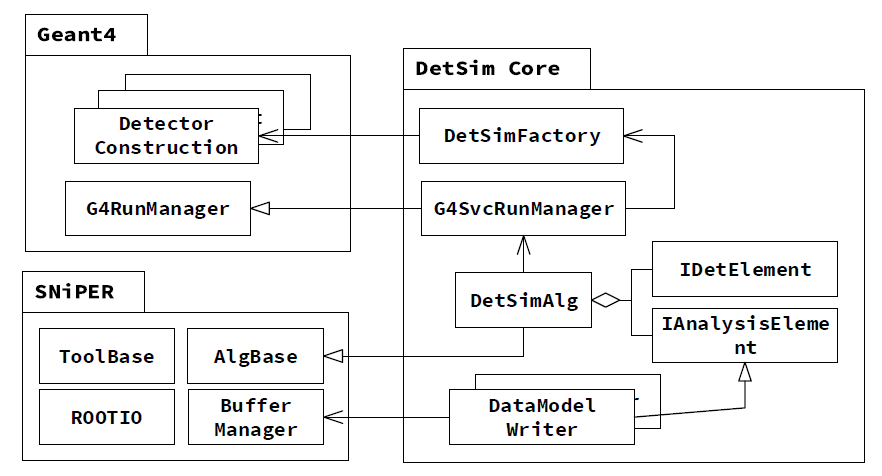}\hspace{2pc}%
\caption{\label{fig:12}Design of detector simulation framework}
\end{center}
\end{figure}

A serial detector simulation framework \cite{Lin:2017usg} is implemented to integrate SNiPER and Geant4 \cite{Agostinelli:2002hh, Allison:2006ve}. Figure~\ref{fig:12} shows the core class diagram. It is lightweight with only a few classes in the design. Simulation software can be easily migrated from standalone application into simulation framework.
However, simulation of a large detector is not only time consuming but also memory and I/Os.
After construction of the full geometry, simulation takes about 700~MB in physical memory. More memory can be occupied during event loop. During simulating of cosmic ray muons, millions of optical photons are produced and propagated. It's a challenge to simulate so many optical photons. A parallelized simulation software is necessary for Monte-Carlo production in the future. Both underlying framework and simulation libraries are required to be thread safe and able to support parallel computing.

\begin{figure}[h]
\begin{center}
\includegraphics[width=20pc]{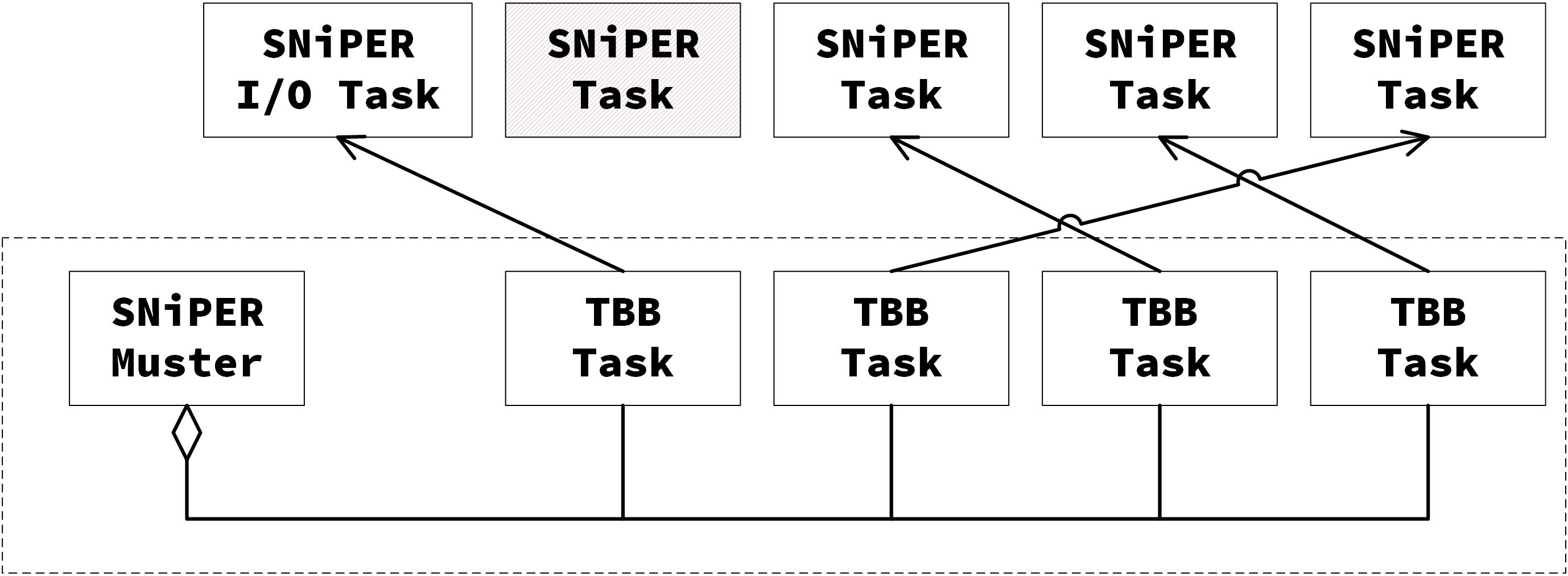}\hspace{2pc}%
\caption{\label{fig:13}SNiPER Muster}
\end{center}
\end{figure}

In the SNiPER framework, a parallelized solution based on Intel Threading Building Blocks (TBB) is proposed. Intel TBB supports task-based programming, which simplifies the thinking of parallel computing.
In SNiPER, {\tt Task} is an important component, which is like a lightweight application manager. In sequential mode, a {\tt Task} is configured by user and then invokes registered algorithms. To support parallel computing, SNiPER Muster (Multiple SNiPER Task Scheduler) is designed and implemented to integrate SNiPER {\tt Task} and Intel TBB. Instead of creating a single instance of {\tt Task}, multiple instances of {\tt Task} are created. However, SNiPER Muster does not execute these {\tt Task}s directly. It will create corresponding TBB-based workers to execute the {\tt Task}. For better performance, each worker is created only once and then binds to one thread. Each worker pulls a {\tt Task} and execute one event. If there are several workers, {\tt Task}s are dispatched to different workers dynamically. As shown in Figure~\ref{fig:13}, an I/O {\tt Task} and three regular {\tt Task}s are running while a regular {\tt Task} is inactive. If the I/O {\tt Task} is done, the inactive {\tt Task} will continue to run.

Starting from version 10.0, Geant4 supports multi-threaded simulation \cite{Allison:2016lfl}. Events are simulated in different threads. With the evolution of Geant4, it is possible to run simulation with Intel TBB, MPI and so on \cite{Dotti:2016ors}. A parallelized simulation framework is developed to integrate the latest Geant4 and SNiPER Muster, which takes full advantages of parallel computing. Based on SNiPER Muster, the simulation framework is decoupled with underlying Intel TBB, and becomes more flexible and simple.

\section{SNiPER Muster based simulation framework}

\begin{figure}[h]
\begin{center}
\includegraphics[width=20pc]{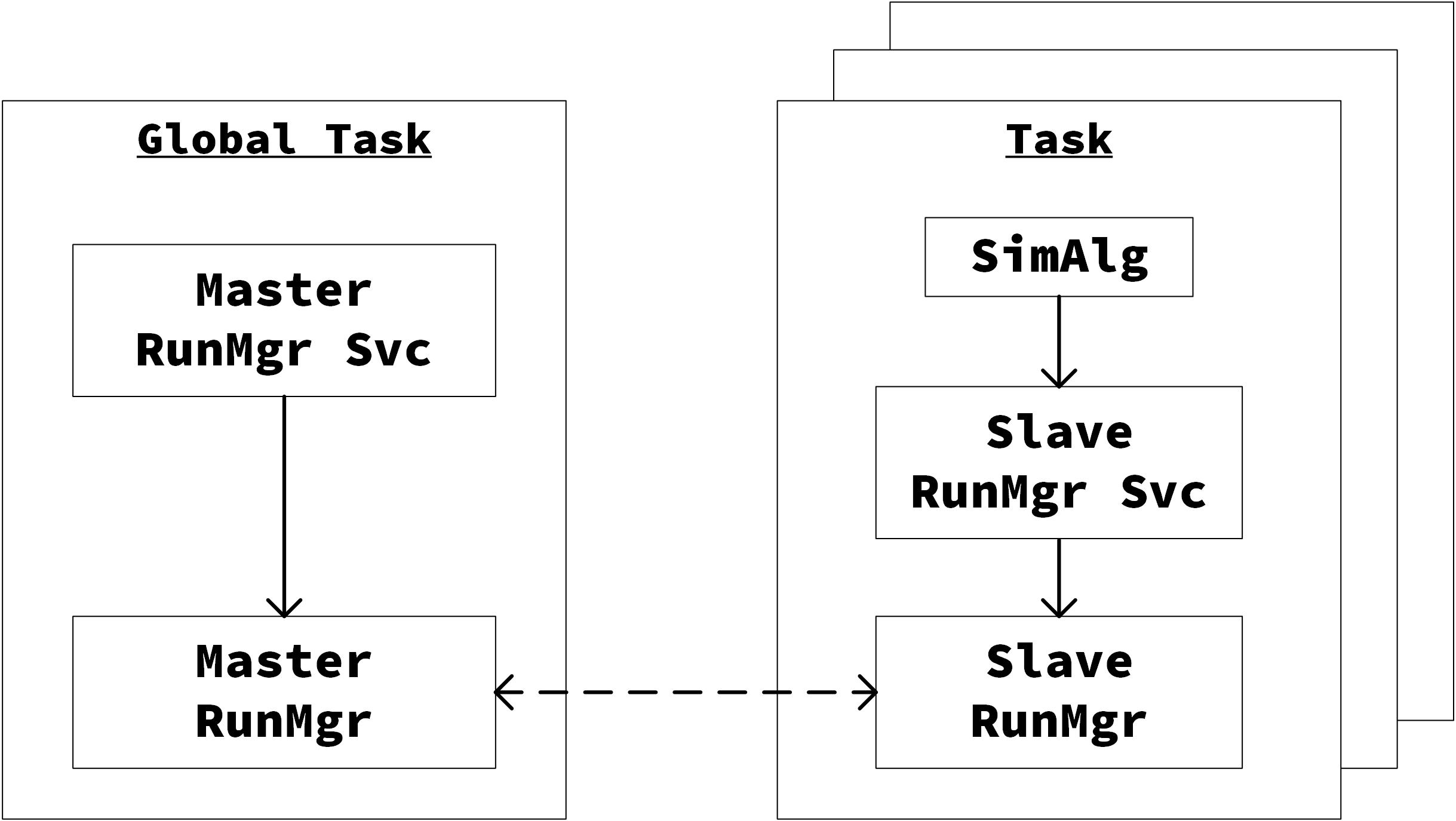}\hspace{2pc}%
\caption{\label{fig:14}Design of parallelized simulation framework}
\end{center}
\end{figure}

The parallelized simulation framework is developed based on SNiPER Muster. Several new classes are introduced, compared with the serial simulation framework. The event loop is still controlled by SNiPER instead of Geant4, however, the control flow is a bit different from serial simulation framework. As shown in Figure~\ref{fig:14}, the simulation framework consists of a global task and several worker tasks. The global task is in charge of initialization of detector geometry and physics list. A customized master run manager is derived from Geant4's {\tt G4MTRunManager} to take control of simulation. A corresponding service is created, so that SNiPER can initialize run manager automatically.

After the initialization of global task, SNiPER Muster starts the worker tasks. Each worker task manages a simulation algorithm and a slave run manager service. The simulation algorithm can invoke the slave run manager via the corresponding service. The slave run manager is derived from Geant4's {\tt G4WorkerRunManager}, which breaks the original event loop and simulates one event each call. Instead of executing fixed number of events in each worker task, the execution of worker task is controlled by SNiPER Muster. If there are events to be simulated, the worker task will invoke the simulation algorithm. Otherwise, the worker task will be stopped. Hence, this event dispatching procedure can make full usage of CPUs.

\section{Performance measurements}
In the performance measurements, 1000 events of single $\gamma$s with energy of 2.2~MeV are generated at detector center. All are done at a blade server with Intel Xeon CPU E5-2680 v3 @ 2.5~GHz and 64~GB of memory. The operating system is Scientific Linux 6.5 with GCC 4.9.4. The Geant4 is 10.03.p01, built with {\tt global-dynamic} TLS (Thread Local Storage)  model. The physics list {\tt G4EmStandardPhysics} and {\tt G4OpticalPhysics} are used in simulation. The simulation time are measured three times for each case. In order to eliminate I/O interference, the events are not saved into ROOT files. 

\begin{figure}[h]
\begin{center}
\includegraphics[width=20pc]{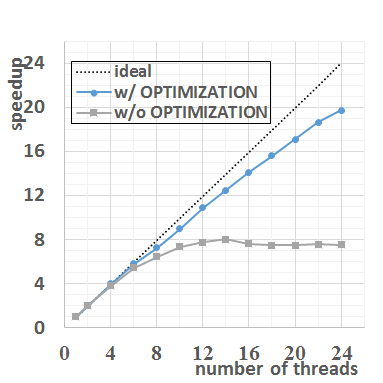}\hspace{2pc}%
\caption{\label{fig:15}Speedup versus number of threads}
\end{center}
\end{figure}

Figure~\ref{fig:15} shows speedup ratio versus number of threads. The dotted line represents the ideal result. The gray line with full square represents the result without any optimization. Up to four threads, the speedup is good. However, when increasing the number of threads to 24, the speedup ratio is stuck around 8. 

Intel VTune Amplifier profiler is used to analyze this performance issue. We find that the hotspot comes from a mutex used in Geant4's {\tt G4MaterialPropertiesTable}. This mutex is used when each optical photon calculates its group velocity. The problem is that there are a lot of optical photons in JUNO simulation, so this mutex is accessed frequently in order to get the velocity. When there are multiple threads accessing the mutex, the performance issue happens. The optimization is moving this part into initialization stage. When refractive index is set, the group velocity is precalculated during the initialization.
As shown in Figure~\ref{fig:15}, the blue line with full circle represents the result with optimization. After this optimization, a better performance is achieved. 

\section{Conclusions}
In this proceeding, we show how JUNO simulation software is parallelized based on SNiPER Muster. By introducing several new classes, Geant4 10 is integrated into SNiPER Muster. The global task initializes detector geometry and physics processes while the worker task is invoked by SNiPER Muster to simulate one event. Events are dispatched dynamically to different workers, which improves CPU utilization. The software performance studies show that there are performance issues due to a mutex in material properties table. After optimization, simulation software achieves a linear speedup, which fulfills requirements.

To speedup events such as cosmic ray muons, we are investigating how to use track-level parallelism in simulation framework. The idea is using MPI to dispatch tracks from master node to different worker nodes. In the future, MPI and TBB can be applied simultaneously in different levels of parallelized simulation software in order to speedup simulation.

\section*{Acknowledgments}
This work is supported by Joint Large-Scale Scientific Facility Funds of the NSFC and CAS (U1532258), the Strategic Priority Research Program of the Chinese Academy of Sciences, Grant No. XDA10010900, National Natural Science Foundation of China (11575224, 11405279, 11675275).

\section*{References}
\bibliography{iopart-num}

\end{document}